# RESONANCES IN THE HEAVY SYMMETRICAL TOP WITH VIBRATING PIVOT

**Shayak Bhattacharjee**

Department of Physics,
Indian Institute of Technology Kanpur,
NH-91, Kalyanpur,
Kanpur – 208016,
Uttar Pradesh, India.

\* \* \* \* \*

## CLASSIFICATION

## ABSTRACT

In this work we consider a heavy symmetrical top whose pivot is subjected to small amplitude, high frequency vibrations in the vertical direction. For analytical simplicity we confine ourselves to positions of the top close to the vertical. We first apply the slow-fast separation method originally devised by Kapitza and Landau for analysing the vibrational stabilization of an inverted pendulum. This analysis yields the slow precession frequency but we see that the equations become undefined at a particular value of the vibration frequency. This breakdown is seen to correspond to a resonance and we use Euler's equations to write down the solution at the resonance. For vibration in the horizontal direction there is a resonance at more or less the same frequency as before but the dynamics at the resonance is different from the former case.

\* \* \* \* \*

# I. INTRODUCTION

It is well known that if the base of an inverted pendulum is vibrated in the vertical direction with small amplitude and high frequency, then the pendulum seems to defy gravity and the inverted vertical in fact becomes a stable equilibrium position. The mathematical details of this stabilization were first proposed by Kapitza [1] and were systematized by Landau [2]. Since these original publications, a large number of studies [3]-[5] of the inverted pendulum have been carried out. Some of these propose physical explanations of the apparently bizarre phenomenon, while others work out extensions and variations.

A second method of stabilization of apparently unstable bodies is rotation, which is what happens in the heavy symmetrical top. The top's speedy rotation about its axis of symmetry prevents it from crashing to the ground under the effect of gravity but instead causes it to precess in a narrow cone about the (inverted of course) vertical position. An enormous number of studies have gone into understanding this stabilization in greater detail – the more comprehensive ones are [6]-[8].

Studies of a combination of these two phenomena are surprisingly rare though. An analysis using the KAM technique may be found in [9] and a related study [10]. The mathematical complexity of the work results in a complete but somewhat recondite treatment. In this work, we apply an intuitive approach. Quite surprisingly, this approach indicates a resonance which the previous two references do not mention explicitly. We get rid of some of the non-linearity by assuming that the top is always close to the vertical. In practice, tops generally remain confined within a small angle and even at 30 degrees, $\sin\theta$ can be replaced by $\theta$ with an error of only 5 percent. We start out by applying Landau's method to the top. We note that a slow-fast separation treatment has also been performed for a top by Berry et. al. [11]. First we briefly summarise Landau's method in its original context.

Consider a simple pendulum with a bob of mass $m$ attached to a massless rigid rod of length $l$. The base of the pendulum is vibrated in the vertical direction with an amplitude $f$ and a frequency $\omega$. Here $\omega$ is significantly greater than the natural frequency $g/l$. Letting $\theta$ be the angular displacement of the bob from the vertically upward position and linearizing, the equation of motion of the bob is found to be

$$l\ddot{\theta} - \left(g + \omega^2 f \cos\omega t\right)\theta = 0 \quad . \tag{1}$$

Solution is achieved by expressing $\theta$ as the sum of a slow varying (frequency of the order of $g/l$) component $\theta_s$ and a fast varying component (frequency approximately $\omega$) $\theta_f$. (1) then reduces to

$$\ddot{\theta}_s + \ddot{\theta}_f - \frac{g}{l}\left(\theta_s + \theta_f\right) = \frac{1}{l}\left(\omega^2 f \cos\omega t\right)\left(\theta_s + \theta_f\right) \quad . \tag{2}$$

Since the slow and fast varying parts of (2) must separately be equal, we have the following two equations :

$$\ddot{\theta}_s - \frac{g}{l}\theta_s = \frac{\omega^2 f}{l}\theta_f \cos\omega t \quad , \tag{3a}$$

$$\ddot{\theta}_f - \frac{g}{l}\theta_f = \frac{\omega^2 f}{l}\theta_s \cos\omega t \quad . \tag{3b}$$

(3b) can be integrated over a time scale corresponding to several periods of $\theta_f$, in which $\theta_s$ does not undergo any appreciable change. Hence we can assume the term $\theta_s$ on its right hand side (RHS) to be a constant while doing this round of integration. Neglecting $g/l$ in comparison with $\omega^2$, we get

$$\theta_f = -\frac{f}{l}\theta_s \cos\omega t \quad . \tag{4}$$

We substitute (4) into (3a) and average over a time scale intermediate between $\omega^{-1}$ and $(l/g)^{1/2}$. Over this averaging period $<\theta_s>$ may still be assumed equal to $\theta_s$ as it varies slowly. On the other hand the cosine squared averages out to 1/2 and we are left with the equation

$$\ddot{\theta}_s + \left(\frac{f^2\omega^2}{2} - \frac{g}{l}\right)\theta_s = 0 \quad . \tag{5}$$

$\theta_s=0$ is clearly a stable fixed point if the quantity in the parentheses of the left hand side (LHS) of (5) exceeds zero i.e. if

$$f^2\omega^2 > 2gl \quad , \tag{6}$$

a conclusion which is in agreement with experimental findings.

## II. THE TOP IN LANDAU'S METHOD

We now consider a heavy symmetric top whose pivot is subjected to high frequency, small amplitude vibrations in the vertical direction. We use the Euler angles $\theta$, $\varphi$ and $\psi$ for this problem. The notation is pretty standard (see for instance Goldstein [12]) so we explain it in short. $\theta$ is the angle made by the symmetry axis (or figure axis) of the top from the vertical. Its derivative thus indicates the nutation. $\varphi$ is an angle in the horizontal plane and its time derivative gives the rate of precession. $\psi$ is an angle in the plane perpendicular to the figure axis and its derivative is the spin rate of the top about its own figure axis. A diagram is given in Sec. III where it is of greater relevance as a geometrical argument is used in that section. The moment of inertia of the top about its figure axis is $I_o$ and about the other two principal axes is $I$. In terms of these principal axes the dynamics of the top can be written as in (7). This derivation can be found in any text book so we do not present it in detail. We just mention that there are two main approaches to the derivation – (a) a direct application of $\text{torque} = d\mathbf{L}/dt$ as in the classic by Morin [13] and (b) an application of Euler's equations in a frame sharing part of the top's angular velocity as in Meriam's standard engineering text [14]. Moreover, as in the pendulum case, we make the replacement $g \rightarrow g + \omega^2 f \cos\omega t$ where $f$ and $\omega$ are the amplitude and frequency of the external vibratory motion. Letting $h$ denote the length of the rod from the pivot to the centre of mass of the top, we have

$$\frac{d}{dt}(\dot{\psi} + \dot{\varphi}\cos\theta) = 0 \Rightarrow \dot{\psi} + \dot{\varphi}\cos\theta = \omega_o = \text{const.} \quad , \tag{7a}$$

$$I\ddot{\varphi}\sin\theta + \dot{\theta}(2I\dot{\varphi}\cos\theta - I_o\omega_o) = 0 \quad , \tag{7b}$$

$$I\ddot{\theta} - \sin\theta(I\dot{\varphi}^2\cos\theta - I_o\omega_o\dot{\varphi}) = Mh\sin\theta(g + \omega^2 f \cos\omega t) \quad . \tag{7c}$$

Two assumptions are used to simplify the modeling. One is that the angle $\theta$ is small. This is physically realistic as large $\theta$ positions are rarely stable configurations for practical tops. Second is that the rate of precession is very slow compared to the rpm of the top. In other words, $\dot{\varphi} \ll \dot{\psi} \Rightarrow \omega_o = \dot{\psi} = \Omega$, where we have used $\Omega$ to denote the rpm of the top which remains approximately constant in time by (7a).

We write $\theta=\theta_s+\theta_f$ where $\theta_f$ varies rapidly (at frequency $\omega$) and $\theta_s$ is slow varying, and analogously let $\varphi=\varphi_s+\varphi_f$. Then the slow parts of (7b) and (7c) are

$$I(\theta_s\ddot{\varphi}_s + \theta_f\ddot{\varphi}_f) - I_o\Omega\dot{\theta}_s = 0 \quad , \tag{8a}$$

$$I\ddot{\theta}_s + I_o\Omega(\theta_s\dot{\varphi}_s + \theta_f\dot{\varphi}_f) = Mh(\theta_s g + \theta_f\omega^2 f \cos\omega t) \quad . \tag{8b}$$

Similarly the fast parts of these equations are

$$I(\theta_s\ddot{\varphi}_f + \theta_f\ddot{\varphi}_s) - I_o\Omega\dot{\theta}_f = 0 \quad , \tag{9a}$$

$$I\ddot{\theta}_f + I_o\Omega(\theta_s\dot{\varphi}_f + \theta_f\dot{\varphi}_s) = Mh(\theta_f g + \theta_s\omega^2 f \cos\omega t) \quad . \tag{9b}$$

The fast equations admit the following solutions :

$$\theta_f = \frac{IMh\omega^3 f}{I_o^2\omega\Omega^2 - I^2\omega^3}\theta_s \cos\omega t \quad , \tag{10a}$$

$$\varphi_f = \frac{I_o Mh\omega^2 \Omega f}{I_o^2\omega\Omega^2 - I^2\omega^3}\sin\omega t \quad . \tag{10b}$$

Substituting these into (8a) and (8b) and averaging over a few cycles, we get the resultant equations for the slow variables as

$$I\theta_s \ddot{\varphi}_s - I_o\Omega\dot{\theta}_s = 0 \quad , \tag{11a}$$

$$I\ddot{\theta}_s + \theta_s \left[ I_o\Omega\dot{\varphi}_s - Mgl - \frac{IM^2h^2\omega^5 f^2}{2\omega(I_o^2\Omega^2 - I^2\omega^2)} + \frac{I_o^2 IM^2h^2\omega^6\Omega^2 f^2}{2\omega^2 (I_o^2\Omega^2 - I^2\omega^2)^2} \right] = 0 \quad . \tag{11b}$$

We will use these equations to obtain corrections on the steady state precession frequency. In steady state, both the terms on the left hand side (LHS) of (11a) are zero and d$\varphi_s$/d$t$ is obtained by setting the coefficient of $\theta_s$ in (11b) equal to zero. For the case of an ordinary top with no vibration, we recover $\dot{\varphi}_s = Mgl/I_o\Omega$ as expected. For the vibrating top, the most surprising feature of (11) is undoubtedly the pole at $I\omega = I_o\Omega$. Clearly, the analysis breaks down in this region and a new approach will be required. Before that, we look at (11) when $\omega \gg \Omega$, a limit where the analysis is clearly valid. After dropping all terms of size $\Omega/\omega$ and smaller, we get

$$\dot{\varphi}_s = \frac{1}{I_o\Omega}\left( Mgl - \frac{M^2 h^2 \omega^2 f^2}{2I} \right) \quad . \tag{12}$$

The negative value indicates that the direction of precession tends to reverse as the product $\omega f$ increases. This can be interpreted in terms of the inverted pendulum. For the pendulum, as $\omega f$ increases, the inverted position changes from an unstable to a stable one, as if gravity reverses direction to point along the inverted vertical. This effect is seen in the top. The "reversed gravity" is responsible for the reversed direction of precession.

We now work in the region where the preceding analysis is not valid.

## III. EULER EQUATIONS

We are very familiar with Euler's equations of rigid body rotation i.e.

$$I_1\dot{\omega}_1 + (I_3 - I_2)\omega_3\omega_2 = \Gamma_1 \quad , \tag{13}$$

and its cyclic permutations but to avoid mistakes we give a brief note of how this equation is derived. It is obtained by transformation from a body fixed frame (the principal basis) i.e. a rotating or non-inertial frame to an inertial frame, which, at any instant of time, is aligned with the body frame. For this one has to use Coriolis' theorem where the subscript aligned refers to the inertial frame which is instantaneously aligned with the body axes. The theorem runs as

$$\left(\frac{d\mathbf{L}}{dt}\right)_{\text{aligned}} = \left(\frac{d\mathbf{L}}{dt}\right)_{\text{body}} + \boldsymbol{\omega} \times \mathbf{L} \quad , \tag{14}$$

and we now employ the fact that in the principal basis, $\mathbf{L} = (I_1\omega_1, I_2\omega_2, I_3\omega_3)$. We have already mentioned that one standard derivation of the dynamical equations employs Euler's equations in a frame sharing part of the top's angular velocity; here we extend that approach to work in a frame sharing the top's entire angular velocity. Subsequent to the Eulerian extraction from the non-inertial to the aligned inertial frame, we can use geometric axis transformations to obtain the angular velocity components with respect to a fixed set of lab frame axes. Since our arguments will be geometrical in character we now present detailed diagrams of the various reference frames used, Figs. 1-2.

Here we have defined the *d-q-o* frame (*) as the one fixed to the body. Since the top is symmetric, it follows that $\omega_o$ will be conserved, and the remaining Euler equations may be written immediately as

$$I\dot\omega_d + (I_o - I)\omega_o\omega_q = \Gamma_d \quad , \tag{15a}$$

$$I\dot\omega_q + (I - I_o)\omega_o\omega_d = \Gamma_q \quad . \tag{15b}$$

It remains to express the *d-q-o* components of $\boldsymbol{\omega}$ and $\boldsymbol{\Gamma}$ in terms of components which are physically observable. In this connection we quote the transformation matrices from the inertial *x-y-z* to the *a-b-c* and the *d-q-o* frames. We have

$$\begin{bmatrix} \hat{\mathbf{a}} \\ \hat{\mathbf{b}} \\ \hat{\mathbf{c}} \end{bmatrix} = \begin{bmatrix} \cos\varphi & \sin\varphi & 0 \\ -\cos\theta\sin\varphi & \cos\theta\cos\varphi & -\sin\theta \\ -\sin\theta\sin\varphi & \sin\theta\cos\varphi & \cos\theta \end{bmatrix} \begin{bmatrix} \hat{\mathbf{x}} \\ \hat{\mathbf{y}} \\ \hat{\mathbf{z}} \end{bmatrix} \quad , \tag{16}$$

and

$$\begin{bmatrix} \hat{\mathbf{d}} \\ \hat{\mathbf{q}} \\ \hat{\mathbf{o}} \end{bmatrix} = \begin{bmatrix} \cos\psi & \sin\psi & 0 \\ -\sin\psi & \cos\psi & 0 \\ 0 & 0 & 1 \end{bmatrix} \begin{bmatrix} \hat{\mathbf{a}} \\ \hat{\mathbf{b}} \\ \hat{\mathbf{c}} \end{bmatrix} \quad . \tag{17}$$

Now geometrical considerations (Fig. 1) yield that $\dot\theta$ is along the *a* axis, $\dot\varphi$ along the *z* axis and $\dot\psi$ along the *o* axis. Using the matrices in (16) and (17) we can calculate their contributions to $\omega_d$, $\omega_q$ and $\omega_o$. Adding all these we get the relations

$$\begin{bmatrix} \omega_d \\ \omega_q \\ \omega_o \end{bmatrix} = \begin{bmatrix} \cos\psi & \theta\sin\psi & 0 \\ -\sin\psi & \theta\cos\psi & 0 \\ 0 & 1 & 1 \end{bmatrix} \begin{bmatrix} \dot\theta \\ \dot\varphi \\ \dot\psi \end{bmatrix} \quad , \tag{18}$$

where $\theta$ has been treated as small. The torque of gravity is along the *a* axis and so is the torque of the vibrations. In this section we are going to neglect gravity and focus only on the vibrations. Hence we essentially treat the free symmetric top with vibrating pivot. Now from the pivot frame the centre of mass (CM) will be experiencing a pseudo force which will vary sinusoidally in time.

---

(*) The labels *a-b-c* and *d-q-o* are taken from engineering convention and have been used to highlight the differences between the various reference frames. *d* stands for "direct", *q* for "quadrature" and *o* for zero-sequence, a terminology perhaps confusing to the present reader who can imagine the letter to stand for "orthogonal".

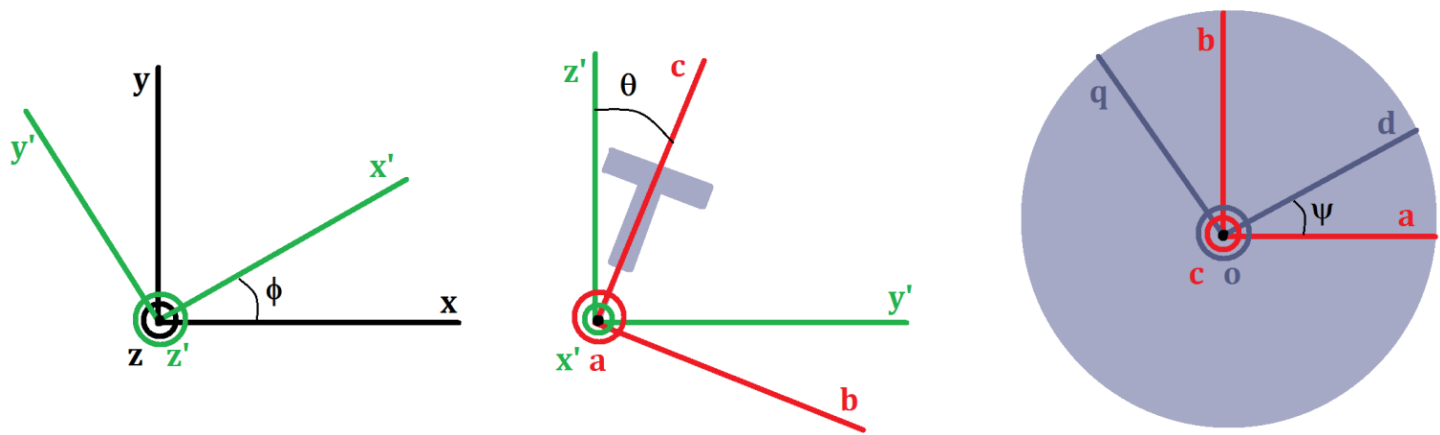

Figure 1 : *Orthographic views of orthogonal transformations. In each view the axis of rotation comes out of the plane of the paper and is shown as a circle in the relevant colour. The x-y-z system is fixed in the lab frame with gravity along the –z direction. The origin is at the top's pivot. The first rotation (left panel) is through angle φ about the z axis to form the x'-y'-z' axes (these axes are useless in the analysis that follows). The next step (middle panel) consists of a rotation through angle θ about the x' axis and the set of axes thus formed is denoted by a-b-c. The axis of symmetry of the top lies along the c axis and the top can be seen in profile view. The final step (right panel) is a rotation through angle ψ about the c axis to get the d-q-o axes. The coloured background highlights the fact that the rotation takes place in the plane of the top's spin. The slate grey colour has been consistently used to indicate the top in this and the next figure; moreover the d q and o axes are shown in a darker shade of the same colour to highlight the fact that they are fixed in the top's frame. We note that a-b-c and d-q-o are both principal axes for the top – however the latter frame carries the entire angular velocity of the top while the former has only a part of it.*

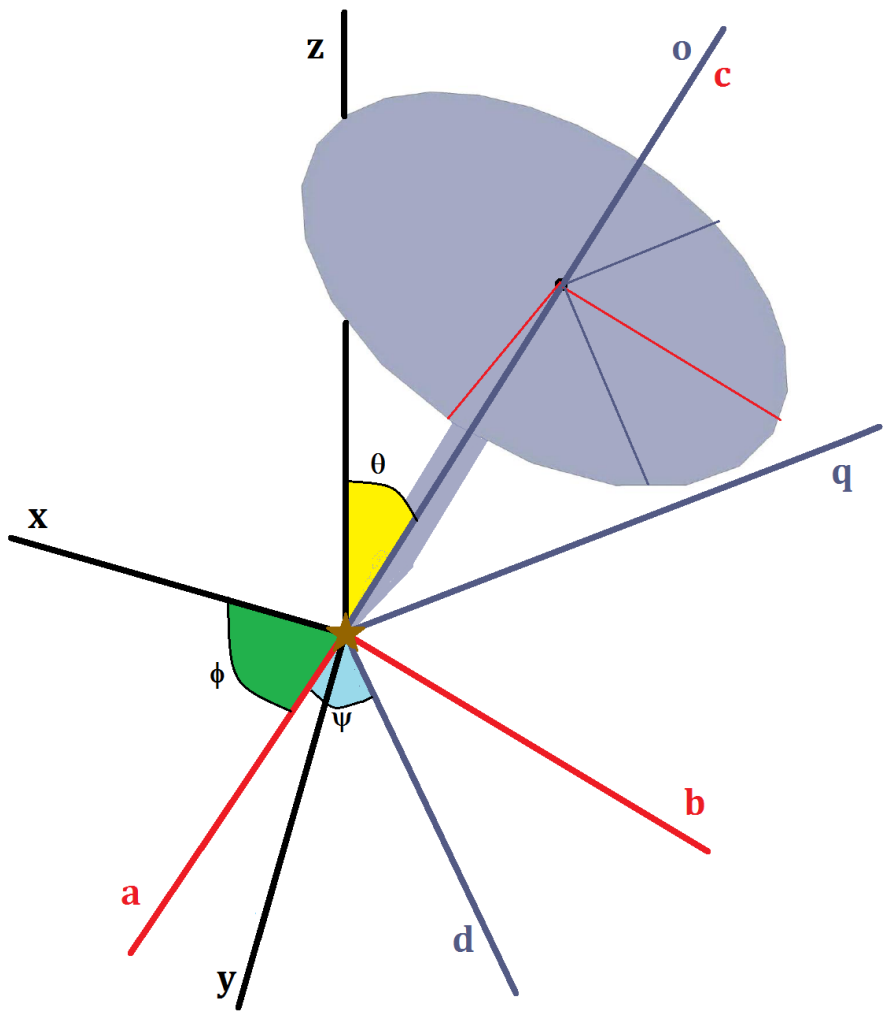

Figure 2 : *Three-dimensional view of the top. The axes and angles mentioned in Fig. 1 can be seen clearly. We should remember that the rotations through φ and ψ are not coplanar. The scheme of rotation follows the "x-convention" mentioned in Goldstein [12].*

We now introduce the space phasor method which is used to reduce the order of systems of real differential equations by constructing complex variables (phasors). Two independent variables are combined into one phasor by adding *j* (imaginary unit) times the second variable to the first; the equations are then written in terms of these phasors. This formulation often expresses the system dynamics in a more elegant way than the original formulation in terms of purely real quantities – applications include the Foucault pendulum [15] and the study of electrical machines [16],[17]. In such a calculation we must keep in mind that the imaginary parts of the phasors are as significant as the real parts. In particular, a loose representation of cos$\omega t$ as $e^{j\omega t}$ is not permissible. Here we form the phasor $\underline{\omega}$ by combining the *d* and *q* components :

$$\underline{\omega} = \omega_d + j\omega_q \quad . \tag{19}$$

Likewise we define the $\underline{\Gamma}$ phasor $\Gamma_d + j\Gamma_q$. Now we substitute the phasor notation and the result of (19) into (15) to obtain a single equation in terms of the phasors : The torque of this force will be the cross product with the displacement vector of the CM and hence will amount to

$$\Gamma_a = M\omega^2 fh\theta \cos\omega t \quad , \tag{20}$$

on which (17) must be applied to obtain the *d*, *q* and *o* components.

$$\frac{d}{dt}\underline{\omega} + j\frac{(I-I_o)\omega_o}{I}\underline{\omega} = \frac{M\omega^2 fh\theta}{I}\cos\omega t \underline{1} e^{-j\psi} \quad . \tag{21}$$

The $\underline{1}$ on the right hand side (RHS) of the above is just the unit phasor $1+j0$, a technical construct to maintain consistency on both sides of the equation (both sides become phasors). Otherwise the coefficient of the exponential must be indicated as a phasor, which is cumbersome notation. This is the equation we are going to solve. Before solving we introduce a simplification. (21) is difficult to handle as $\psi$ which occurs in the exponent is a function of *t*. However, for a fast top we can assume that the change in spin speed is small compared with the speed itself. This is physically plausible. Practical gyroscopes have large moments of inertia and high spin speeds so a large change in rpm implies a huge change in angular momentum, which is not feasible. Under this assumption we make a substitution $\psi = \Omega t$ in (21), where $\Omega$ is a constant. Further we have from (18) and (20),

$$\underline{\omega} = (\dot{\theta} + j\theta\dot{\varphi})\underline{1} e^{-j\Omega t} \quad . \tag{22}$$

We let $\nu = (I-I_o)\omega_o / I$ and $\lambda = M\omega^2 fh$. Solving (21) means separately finding the homogeneous and particular solutions. We first treat the homogeneous solutions which correspond to the free symmetric top. We have the solution

$$\underline{\omega} = \underline{\omega}_{00} e^{-j\nu t} \quad , \tag{23}$$

where $\underline{\omega}_{00}$ depends on the initial conditions. The double cipher avoids any potential confusion with the *o* used as axis label. Equating the RHS of (22) and (23) we have the solution

$$(\dot{\theta} + j\theta\dot{\varphi})\underline{1} = \underline{\omega}_{00} e^{j(\Omega-\nu)t} \quad . \tag{24}$$

Since $\theta$ and $\varphi$ are real, the real and imaginary parts of the left hand side (LHS) are obvious, and they must separately equal the corresponding parts on the RHS. As an example, we obtain the criterion for precession without nutation. For that we must have $\dot{\theta}=0$ whereby the LHS becomes purely imaginary at all time. Now the coefficient of *j* in the exponential on the RHS must be zero otherwise the real and imaginary parts of the RHS will fluctuate in time. This requirement forces $\Omega=\nu$ which at once determines the precessional frequency $\frac{I_o\Omega}{I-I_o}$, that can be shown with a little manipulation to agree with the standard result $I_o\omega_o/I$. This is a necessary condition but is certainly not sufficient for regular precession – there is an additional constraint that $\underline{\omega}_{00}$ must be purely imaginary. In other words, the steady precession state can only be initiated by holding the top at a constant angle $\theta$ and giving it an impulse in the $\varphi$ direction. Nutation, on the other hand, is achieved by giving the free top an initial impulse in the $\theta$ direction. In this case $\underline{\omega}_{00}$ is real and the real part of (24) yields

$$\dot{\theta} = \omega_{00} \cos(\Omega - \nu)t \quad . \tag{25}$$

This describes nutation at frequency $\dfrac{I_o \Omega}{2I - I_o}$; if in addition the initial release angle be zero, we can equate the imaginary parts to get uniform precession at the same frequency. These calculations highlight the advantage of our formalism : it gives a direct link between the final state and the initial conditions. This method can also be used profitably for an examination of the normal behaviour of the heavy symmetric top with or without friction at the pivot. This calculation has been indicated extensively in [18] and we now specialize to the case which we have at hand.

The particular solution of (21) where we must be careful to write the cosine as the sum of a positive and negative complex exponential is

$$\underline{\omega} = \mathbf{1}\frac{j\lambda\theta}{2I}\left(-\frac{e^{j(\omega-\Omega)t}}{\omega-\Omega+\nu} + \frac{e^{-j(\omega+\Omega)t}}{\omega+\Omega-\nu}\right) \quad . \tag{26}$$

Clearly a resonance is possible if either of the two denominators in the above expression vanishes. Of these the first term is seen to vanish when

$$\omega = \Omega - \nu = \frac{I_o}{I}\Omega - \frac{I - I_o}{I}\omega_p \quad , \tag{27}$$

where $\omega_p$ denotes the frequency of precession. If the vibrating top is released from rest without precession, we have $\omega_p=0$ and thus recover the condition at which the Landau analysis fails. If on the other hand the top has a precessional motion, then the resonance frequency changes on account of it. Vanishing of the second denominator in the RHS of (26) calls for a physically unrealistic criterion. Before determining the character of the resonance, we work in the off-resonance region. We have from the real parts of (22) and (26) :

$$\dot{\theta} = \frac{\lambda\theta}{I}\left(\frac{\omega}{\omega^2 - (\Omega - \nu)^2}\right)\sin\omega t \quad . \tag{28}$$

Its integral is the exponential of a sine : for small $\lambda$ the formula reduces to

$$\theta = \theta_0\left(1 - \frac{\lambda}{I(\omega^2 - (\Omega - \nu)^2)}\cos\omega t\right) \quad , \tag{29}$$

which consists of a small oscillatory part about a constant position. Likewise, equating the imaginary parts of (22) and (26) yields the formula for the precession as

$$\dot{\varphi} = \frac{\lambda(\nu - \Omega)}{I(\omega^2 - (\Omega - \nu)^2)}\cos\omega t \quad . \tag{30}$$

Conservation of $\omega_o$ will yield the fluctuation in $\dot{\psi}$ correct to first order in $\lambda$.

We now treat the case of resonance. In this case, (21) will admit solutions of the form $te^{j\nu t}$ and retracing the steps leading to (28) we will get something of the form

$$\dot{\theta} = \alpha\theta(t\cos\nu t) \quad, \tag{31}$$

in which we have been slack about the coefficients and the exact phase of the sinusoid; as we shall see we can afford this. The solution will have two terms from integrating the *t* function by parts – the important term is of the form

$$\theta = \beta\exp(\alpha t\sin\nu t) \quad, \tag{32}$$

which blows up for all values of *β*. The argument of the exponential fluctuates periodically with ever increasing amplitude. The exponential itself then comes close to zero when the argument is large negative and becomes huge when the argument is large positive. Thus, the top performs nutation with ever increasing amplitude, and is quickly thrown off the vertical position. We note that the divergence is stronger than the $t\sin\nu t$ behaviour seen in ordinary resonances, as in simple harmonic oscillators. The phenomenon going on here is more similar to the parametric resonance mentioned in [2].

We briefly examine what happens if the vibration occur not in the *z* direction but in the horizontal plane, say in the *x* direction. Resolving the force into *d* and *q* components and taking the cross product with $h\hat{\mathbf{o}}$ yields the following differential equation for the top :

$$\frac{d}{dt}\underline{\omega} + j\nu\underline{\omega} = j\frac{Mh\omega^2 f\,e^{-j\varphi}}{2}\left(e^{j\omega t} + e^{-j\omega t}\right)e^{-j\Omega t} \quad. \tag{33}$$

We note that the motion can be split up into fast and slow parts as in the starting analysis, and the fast parts will have small amplitude. For this reason, the leading order contribution in the RHS will be from the slow part of *φ*, which we assume has the form $\omega_p t$. Performing the substitution we once again get a resonance if either of the frequencies in the RHS equals *ν*; like the previous case, the first term is the one which does so in a physically plausible manner. The resonance condition is thus

$$\nu = \omega - \Omega - \omega_p \Rightarrow \omega - \frac{I_o}{I}\omega_o = 0 \quad. \tag{34}$$

The resonance frequency is independent of the precessional motion of the top. The resonant behaviour is also different. Solving (32) at resonance and separating real and imaginary parts yields the relation

$$\dot{\theta} = \alpha t\cos\nu t \quad, \tag{35}$$

whose solution with dominant term

$$\theta = \beta t\sin\nu t \tag{36}$$

can be seen to be closer to the usual form of the resonance behaviour.

## IV. CONCLUSION

We have thus analysed the behaviour of a heavy symmetric top whose pivot is subjected to vertical and horizontal vibrations at small amplitude and high frequency. The most interesting feature of the dynamics are the resonances where the top is thrown from its vertical position. The different characters of the two resonances are also noteworthy. Away from the resonance, the effect of the vibration manifests as a correction on the precessional frequency and a small-amplitude oscillatory motion at the vibration frequency. In this regime we are in agreement with [9] which has also obtained a correction on the nutation angle of the order of *f/h* – with considerably more exertion than our derivation. These results are commensurate with those of a simple pendulum with vibrating pivot, where the vibration effectively causes the gravitational force to weaken.

We briefly suggest experimental procedures to verify our results. A motorized gyroscope should be used to prevent the rotation speed from decaying in time. At the same time it must be ensured that the motor does not tend to hold the rpm constant as this would be in violation of the dynamics given here. The torque of the motor should be the controlled variable and should equal the damping torque exerted on the top by its bearings. This will best correspond to the condition $\Gamma_o=0$. The top should be mounted on a vibrating table and the pivot should be clamped securely to it. There should be no lateral displacement of the top during the experiment. Also, as the vibration frequency increases and the effective gravity reduces, the top will exhibit a tendency to become "light" on the table and slip; this should be checked. The effect of the resonance is easily observable; the other phenomena may best be studied by attaching a laser pointer to the top and analysing the motion of the spot of light on a screen far away – this will automatically amplify the small vibratory motions.

In conclusion we would like to point out that we have proposed a novel approach which can be applied to a wide variety of rotational dynamics problems, not just the specific case considered in this paper.

\*     \*     \*     \*     \*

# ACKNOWLEDGEMENT

I am grateful to KVPY, Government of India for a generous Fellowship.